\documentclass[]{aastex631}

\begin{document}

\title{A candidate dark matter deficient dwarf galaxy in the Fornax cluster
identified through overluminous star clusters}

\author[0000-0003-2473-0369]{Aaron J.\ Romanowsky}
\affiliation{Department of Physics \& Astronomy, San Jos\'e State University, One Washington Square, San Jose, CA 95192, USA}
\affiliation{Department of Astronomy \& Astrophysics, University of California Santa Cruz, 1156 High Street, Santa Cruz, CA 95064, USA}

\author{Enrique Cabrera}
\affiliation{Department of Physics \& Astronomy, San Jos\'e State University, One Washington Square, San Jose, CA 95192, USA}

\author[0000-0003-0327-3322]{Steven R.\ Janssens}
\affiliation{Centre for Astrophysics and Supercomputing, Swinburne University, John Street, Hawthorn, VIC 3122, Australia}
\affiliation{ARC Centre of Excellence for All Sky Astrophysics in 3 Dimensions (ASTRO 3D), Australia}



\begin{abstract}

Two low surface brightness (LSB) dwarf galaxies were identified recently 
as having little or no dark matter (DM), provoking widespread interest in their formation histories.
These galaxies also host populous systems of star clusters that are on average larger and more luminous than typical globular clusters (GCs).
We report an initial attempt to identify new candidate DM-deficient dwarfs via their unusual GC systems.
Using a large catalog of LSB galaxies from the Dark Energy Survey,
we inspect their Dark Energy Camera Legacy Survey (DECaLS) imaging and identify FCC~224 as a candidate
found on the outskirts of the Fornax cluster.
We analyze the GC system using DECaLS and archival 
{\it Hubble Space Telescope} WFPC2 imaging, and find an apparent population of overluminous GCs.
More detailed follow-up of FCC~224 is in progress.
%This example manuscript is intended to serve as a tutorial and template for
%authors to use when writing their own AAS Journal articles. The manuscript
%includes a history of \aastex\ and includes figure and table examples to illustrate these features. Information on features not explicitly mentioned in the article can be viewed in the manuscript comments or more extensive online
%documentation. Authors are welcome replace the text, tables, figures, and
%bibliography with their own and submit the resulting manuscript to the AAS
%Journals peer review system.  The first lesson in the tutorial is to remind
%authors that the AAS Journals, the Astrophysical Journal (ApJ), the
%Astrophysical Journal Letters (ApJL), the Astronomical Journal (AJ), and
%the Planetary Science Journal (PSJ) all have a 250 word limit for the 
%abstract\footnote{Abstracts for Research Notes of the American Astronomical 
%Society (RNAAS) are limited to 150 words}.  If you exceed this length the
%Editorial office will ask you to shorten it. This abstract has 161 words.

\end{abstract}



\section{Introduction} \label{sec:intro}

%\latex\ \footnote{\url{http://www.latex-project.org/}} is a document markup
%language that is particularly well suited for the publication of
%mathematical and scientific articles \citep{lamport94}. \latex\ was written
%in 1985 by Leslie Lamport who based it on the \TeX\ typesetting language
%which itself was created by Donald E. Knuth in 1978.  In 1988 a suite of
%\latex\ macros were developed to investigate electronic submission and
%publication of AAS Journal articles \citep{1989BAAS...21..780H}.  Shortly
%afterwards, Chris Biemesdefer merged these macros and more into a \latex\
%2.08 style file called \aastex.  These early \aastex\ versions introduced
%many common commands and practices that authors take for granted today.
%Substantial revisions
%were made by Lee Brotzman and Pierre Landau when the package was updated to
%v4.0.  AASTeX v5.0, written in 1995 by Arthur Ogawa, upgraded to \latex\ 2e
%which uses the document class in lieu of a style file.  Other improvements
%to version 5 included hypertext support, landscape deluxetables and
%improved figure support to facilitate electronic submission.  
%\aastex\ v5.2 was released in 2005 and introduced additional graphics
%support plus new mark up to identifier astronomical objects, datasets and
%facilities.

Since ultra-diffuse galaxies (UDGs) were highlighted by \cite{vanDokkum15} as
an abundant class of LSB dwarf in the Coma cluster,
these intriguing systems have attracted widespread 
observational and theoretical attention.
The first dynamical studies 
focused on cases with unusually populous GC systems and found them to have overmassive DM halos
(e.g., \citealt{Beasley16,vanDokkum16}). 
The nearby UDG NGC~1052-DF2 (hereafter DF2) was subsequently targeted for dynamical study because of its GC richness, 
which yielded the unexpected result that the galaxy is apparently devoid of 
DM \citep{vanDokkum18}.
Intense scrutiny of this claim followed, particularly including doubts about the distance, 
which were addressed with a definitive measurement of $\simeq 22$~Mpc using the tip of the red giant branch
\citep{Shen21b}.

A second galaxy in close proximity, NGC~1052-DF4, was also found to be DM deficient \citep{vanDokkum19},
and attention has turned now to theoretical understanding of
the origins of these peculiar systems.
The two leading scenarios are tidal stripping
and a high-speed ``bullet-dwarf'' collision
(e.g., \citealt{Moreno22,Lee24}).

Finding more analogs to DF2 and DF4
in the nearby Universe where they can be studied in detail ($\lesssim$~30 Mpc) will be invaluable for understanding the formation histories of DM-deficient galaxies.
Deep, blind dynamical surveys of quiescent LSB dwarfs are prohibitively costly in telescope time, 
and methods to select likely DM-deficient candidates for follow-up are needed.
One approach is to capitalize on a unique trait shared by DF2 and DF4 other than their DM-deficiency:
populous systems of GCs that are {\it overluminous}, with
$M_V \sim -9$ mag,
four times brighter than the 
peak GC magnitude of $M_V \sim -7.5$ in other galaxies
\citep{Shen21a}.
The GCs in DF2 were bright enough to discern in Sloan Digital Sky Survey imaging
($g \sim 23$ mag), 
leading to focused follow-up observations of this galaxy, 
and they are even more apparent in DECaLS imaging  (Figure~\ref{fig1}{\it a}).
If other LSB dwarfs at $\sim$~20~Mpc distances have similarly bright GCs, these should also be easily identifiable.

\section{Search and Discovery}

%% The "ht!" tells LaTeX to put the figure "here" first, at the "top" next
%% and to override the normal way of calculating a float position
\begin{figure}[ht!]
\includegraphics[width=\textwidth]{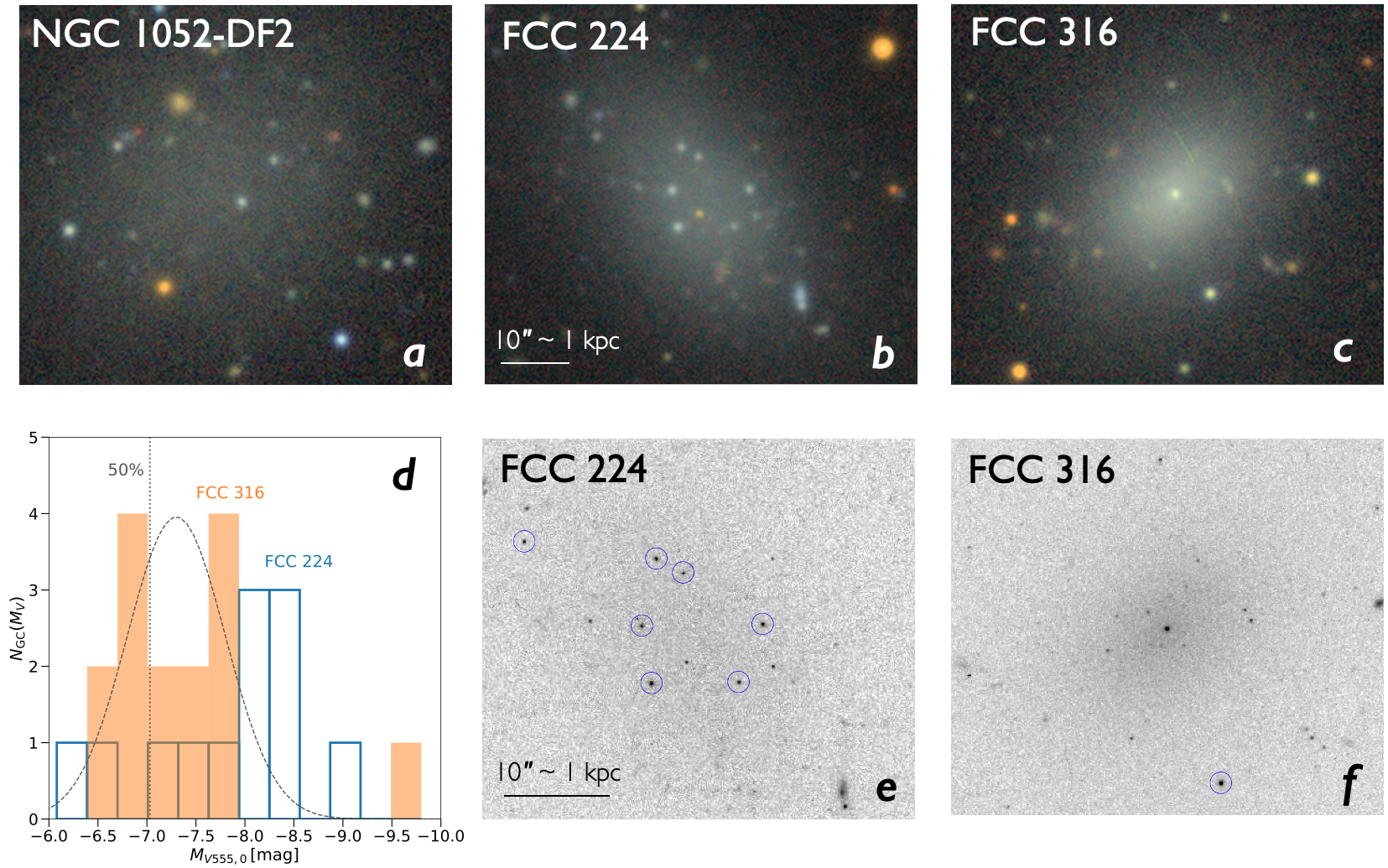}
\caption{Panels ({\it a},{\it b},{\it c}): 
DECaLS DR10 images of NGC~1052-DF2, FCC~224, and FCC~316,
on the same spatial and brightness scales.
Panel ({\it d}): GC luminosity functions of FCC~316 (orange filled histogram),
FCC~224 (blue open histogram), and a standard model (dashed curve).
The 50\% completeness limit is shown as a vertical dotted line.
Panels ({\it e},{\it f}):  {\it HST}/WFPC2 F555W images of FCC~224 and FCC~316.
GC candidates brighter than $M_V \sim -8$ are marked with dark blue circles.
\label{fig1}}
\end{figure}

Our approach here to finding DF2/DF4 analogs is to begin with an existing sample of LSB dwarfs and then identify those that host bright GC candidates.
For rapid, qualitative results, we use visual inspection, as motivated by
the obvious GC-richness of DF2 in the DECaLS image.
The starting point is a catalog of 21,000 LSB galaxies from the Dark Energy Survey 
(\citealt{Tanoglidis21}; catalog version 1).
Only $\sim$~300 of these have reported effective radii large enough to be nearby UDGs
(i.e., $R_{\rm e} \geq$~1.5 kpc means $R_{\rm e} > $10--15~arcsec for distances of 20--30~Mpc).
Still, we do not rely on the $R_{\rm e}$ measurements, and visually inspect images from the entire catalog -- 
using color thumbnails from DECaLS DR8\footnote{\url{https://www.legacysurvey.org/viewer?layer=ls-dr8}}.  
We find a handful of cases with apparently rich systems of star clusters, 
including DF2 which is part of the catalog,
but only one new candidate for hosting overluminous GCs: FCC~224
(Figure~\ref{fig1}{\it b}; discussed in the next section).
We subsequently inspected DECaLS imaging for the entire Fornax Cluster Catalog (FCC; \citealt{Ferguson89}) and found no additional DF2 analogs.

\section{FCC 224 from DECaLS and {\it HST}/WFPC2 imaging}

FCC~224 is an LSB dwarf located on the outskirts of the Fornax cluster in projection.
Although it has a redshift listed in the NASA Extragalactic Database, 
this is not an actual measurement but can be traced to an assumption in 
\citet{Dabringhausen16} that the galaxy belongs to the Fornax cluster.
Here we assume that the Fornax cluster membership is correct, and that the distance is 20~Mpc
\citep{Blakeslee09},
leaving careful distance analysis to future papers.
Photometric analyses in the literature imply a luminosity in the range $L_V \sim (4$--$8)\times 10^7 L_\odot$ and $R_{\rm e}$ in the range $0.9$--$1.7$~kpc
\citep{Tanoglidis21,Zaritsky22,Paudel23}.
The optical color of the galaxy was found to be consistent with the red sequence, i.e., quiescence, while it differs from DF2 and DF4 in being relatively flattened, with an axis ratio of $b/a \sim 0.65$.

Turning to the star cluster system of FCC~224,
DECaLS aperture photometry implies magnitudes of $M_g \sim -8$ to $-9$~mag for
the visually identified GC candidates, and colors of $g-z \sim 1.0$.
These objects have similar colors to GCs in other Fornax dwarf galaxies \citep{Masters10},
but the magnitudes are brighter than average.
Figure~\ref{fig1}{\it b},{\it c} shows this comparison visually with a DECaLS image of FCC~316,
a typical dwarf elliptical in the Fornax cluster,
where a bright nucleus is readily seen, but any GC candidates are hard to discern.

The only previous study of the FCC~224 star clusters was \citet{Seth04}, who used
{\it HST}/WFPC2 to estimate
$11.7\pm2.6$ clusters in the galaxy (including a correction for contamination
but not for incompleteness).
FCC~224 was not in their subsample of galaxies examined in detail because the distance was not confirmed.
Here we return to the WFCP2 data (GO 7377, observation date 1998 Dec 31),
which consists of two 460~s exposures in F555W (hereafter $V_{555}$),
and one 300~s exposure in F814W (hereafter $I_{814}$).
Figure~\ref{fig1}{\it e} shows the $V_{555}$-band image, where the GC candidates are visible as bright, pointlike objects.
A WFPC2 image of FCC~316 from the same program
(GO~6352, observed on 1996 Oct 25) is shown in panel {\it f}, 
where the GC candidates are now more apparent than in DECaLS, but still appreciably fainter than in FCC~224.

We downloaded reduced Hubble Legacy Archive WFPC2 images of FCC~224 and FCC~316,
and used similar methods
to \citet{Janssens22} to measure magnitudes of compact objects, and to estimate completeness
(50\% complete at $V_{555,0} = 25.6$~AB~mag).
The main difference was to use model rather than empirical point-spread functions.
Our GC selection is based on $<20^{\prime\prime}$ spatial position from the galaxy center,
$V_{555,0} = 22$--$26$~magnitude, and $(V_{555}-I_{814})_0 = 0.4$--$0.8$~color --
where some colors had to be estimated from DECaLS
owing to cosmic ray contamination in the $I_{814}$-band images.

The results are summarized in Figure~\ref{fig1}{\it d} as 
histograms of GC numbers after subtracting off a background distribution obtained at larger distances from the galaxies, using essentially the same selection criteria.
The dashed curve shows a standard GC luminosity function for dwarfs with peak of $M_{V} = -7.3$~mag
and dispersion of $\sigma = 0.5$~mag \citep{Seth04,Villegas10}.
FCC~316 appears consistent with the standard distribution, 
while FCC~224 hosts seven GC candidates that are remarkably bright
($M_V = -8$ to $-9$~mag).
Our quantitative analysis thus supports the visual impression that FCC~224 
hosts unusually luminous GCs.
This result motivated follow-up observations using {\it HST}/WFC3 to study
the GC properties in more detail (GO~17149; PI A.~Romanowsky; Y.~Tang et al., submitted),
while efforts to measure the galaxy's DM content are underway
(M.~L.~Buzzo et al., in preparation).

\begin{acknowledgments}

AJR was supported by National Science Foundation grant AST-2308390.
We thank Yimeng Tang for comments.

\end{acknowledgments}


%\bibliography{sample631}{}
\bibliography{reference}{}
\bibliographystyle{aasjournal}



\end{document}